\UseRawInputEncoding
\documentclass[amsmath,amssymb,prx,superscriptaddress,reprint,showpacs,longbibliography]{revtex4-2}
\usepackage{times}
\usepackage{graphicx}
\usepackage{amsfonts}
\usepackage{amsmath, amsthm, amssymb}
\usepackage{dsfont}
\usepackage{color}
\usepackage{empheq}
\usepackage{standalone}
\usepackage[most]{tcolorbox}
\usepackage{physics}
\usepackage{bm}
\usepackage{amssymb,amsmath,amsfonts,amscd}
\usepackage{mathrsfs}
\usepackage{siunitx}
\usepackage{commath}
\usepackage{graphicx}      
\graphicspath{ {Images/} } 
\usepackage{braket}
\usepackage{bm}
\usepackage{empheq}
\usepackage[most]{tcolorbox}
\usepackage{physics}
\usepackage{placeins}

\usepackage{hyperref} 
\hypersetup{
    colorlinks=true,
    linkcolor=blue,    
    citecolor=blue,    
    urlcolor=blue,     
}

\begin{document}

\title{Insulator-to-Metal Crossover near the Edge of the Superconducting Dome in Nd$_{1-x}$Sr$_x$NiO$_2$}

\author{Yu-Te Hsu}
\email{yute.hsu@ru.nl}
\affiliation{High Field Magnet Laboratory (HFML-EMFL) and Institute for Molecules and Materials, Radboud University, Toernooiveld 7, 6525 ED Nijmegen, Netherlands}

\author{Bai Yang Wang}
\affiliation{Department of Physics, Stanford University, Stanford, CA 94305, United States}
\affiliation{Stanford Institute for Materials and Energy Sciences, SLAC National Accelerator Laboratory, Menlo Park, Stanford, CA 94025, United States}

\author{Maarten Berben}
\affiliation{High Field Magnet Laboratory (HFML-EMFL) and Institute for Molecules and Materials, Radboud University, Toernooiveld 7, 6525 ED Nijmegen, Netherlands}

\author{Danfeng Li}
\affiliation{Stanford Institute for Materials and Energy Sciences, SLAC National Accelerator Laboratory, Menlo Park, Stanford, CA 94025, United States}
\affiliation{Department of Applied Physics, Stanford University, Stanford, CA 94305, United States}
\affiliation{Department of Physics, City University of Hong Kong, Kowloon, Hong Kong, China}

\author{Kyuho Lee}
\affiliation{Department of Physics, Stanford University, Stanford, CA 94305, United States}
\affiliation{Stanford Institute for Materials and Energy Sciences, SLAC National Accelerator Laboratory, Menlo Park, Stanford, CA 94025, United States}

\author{Caitlin Duffy}
\affiliation{High Field Magnet Laboratory (HFML-EMFL) and Institute for Molecules and Materials, Radboud University, Toernooiveld 7, 6525 ED Nijmegen, Netherlands}

\author{Thom Ottenbros}
\affiliation{High Field Magnet Laboratory (HFML-EMFL) and Institute for Molecules and Materials, Radboud University, Toernooiveld 7, 6525 ED Nijmegen, Netherlands}

\author{Woo Jin Kim}
\affiliation{Stanford Institute for Materials and Energy Sciences, SLAC National Accelerator Laboratory, Menlo Park, Stanford, CA 94025, United States}
\affiliation{Department of Applied Physics, Stanford University, Stanford, CA 94305, United States}

\author{Motoki Osada}
\affiliation{Stanford Institute for Materials and Energy Sciences, SLAC National Accelerator Laboratory, Menlo Park, Stanford, CA 94025, United States}
\affiliation{Department of Materials Science and Engineering, Stanford University, Stanford, CA 94305, United States}

\author{Steffen Wiedmann}
\affiliation{High Field Magnet Laboratory (HFML-EMFL) and Institute for Molecules and Materials, Radboud University, Toernooiveld 7, 6525 ED Nijmegen, Netherlands}

\author{Harold Y. Hwang}
\email{hyhwang@stanford.edu}
\affiliation{Stanford Institute for Materials and Energy Sciences, SLAC National Accelerator Laboratory, Menlo Park, Stanford, CA 94025, United States}
\affiliation{Department of Applied Physics, Stanford University, Stanford, CA 94305, United States}

\author{Nigel E. Hussey}
\email{nigel.hussey@ru.nl}
\affiliation{High Field Magnet Laboratory (HFML-EMFL) and Institute for Molecules and Materials, Radboud University, Toernooiveld 7, 6525 ED Nijmegen, Netherlands}
\affiliation{H. H. Wills Physics Laboratory, University of Bristol, Tyndall Avenue, Bristol BS8 1TL, United Kingdom}

\date{\today}

\begin{abstract}
We report a systematic magnetotransport study of superconducting infinite-layer nickelate thin films Nd$_{1-x}$Sr$_x$NiO$_2$ with $0.15 \leq x \leq 0.225$. By suppressing superconductivity with out-of-plane magnetic fields up to 37.5~T, we find that the normal state resistivity of Nd$_{1-x}$Sr$_x$NiO$_2$ is characterized by a crossover from a metallic $T^2$-behavior to an insulating log(1/$T$)-behavior for all $x$ except $x = 0.225$, at which the resistivity is predominantly metallic. The log(1/$T$)-behavior is found to be robust against magnetic fields, inconsistent with scenarios involving localization or Kondo scattering, and points to an anomalous insulating state possibly driven by strong correlations. In the metallic state, we find no evidence for non-Fermi-liquid behavior arising from proximity to a putative quantum critical point located inside the superconducting dome.
\end{abstract}

%\keywords{}%Use showkeys class option if keyword
                              %display desired

\maketitle
%\section*{Introduction}
The realization of superconductivity in infinite-layer nickelates \cite{li2019} has created a new model system for studying unconventional superconductivity \cite{hirsch2019, botana2020, wu2020, been2020, lechermann2020, jiang2020, zhang2020,  sakakibara2020, choi2020, kang2020}. While the defining framework of the high-$T_{\rm c}$ cuprates, namely a 3$d^9$ electronic configuration in square planar geometry, is reproduced in the infinite-layer nickelates, the extent to which the two oxide families can be considered analogous remains an outstanding open question. 

The relative strength of the electron correlations, for one, is currently unclear. Undoped cuprates are antiferromagnetic Mott insulators \cite{lee2006}. Undoped bulk NdNiO$_2$, by contrast, does not order down to 1.7~K \cite{hayward2003} while thin film NdNiO$_2$ displays only weakly insulating behavior below $\approx$ 70~K \cite{li2019}. The re-entrant insulating behavior found in nickelates at high doping \cite{li2020, zeng2020, osada20201, osada2021, zeng2021} further contrasts with the correlated Fermi-liquid ground state found in highly overdoped cuprates \cite{nakamae2003}. Thirdly, various theoretical calculations \cite[among others]{lee2004, hepting2020, botana2020, wu2020, been2020, lechermann20201} have thus far found that the low-energy electronic structure of infinite-layer nickelates contains contributions from both the Ni 3$d_{x^2-y^2}$ and Nd 5$d$-orbitals -- rather than the single-band picture established for the cuprates \cite{lee2006} -- though the role or relevance of the Nd 5$d$-orbitals remains controversial, particularly in superconducting (SC) samples \cite{kitakani2020}. X-ray \cite{hepting2020}, electron energy loss spectroscopy \cite{goodge2020} and Hall coefficient \cite{li2020, zeng2020} studies, however, do seem to indicate the presence of multiple bands.  Lastly, the peak critical temperature $T_{\rm c} \sim$ 10~K in nickelates is comparatively low, raising the question whether the two families even share the same pairing mechanism. Evidently, further investigation of the normal state from which nickelate superconductivity emerges is required to ascertain what commonalities, if any, they share with the cuprates.

Typically, the phase diagram of unconventional superconductors harbours a dome of superconductivity bisected by a transition line to some other order, which competes with superconductivity for the dominant ground state \cite{shibauchi2014}. One strategy for determining the evolution of this transition line inside the superconducting (SC) dome and the influence of its termination (at a putative quantum critical point) on the transport properties is to suppress the superconductivity in a high magnetic field and to study the form of the resistivity down to the lowest possible temperatures \cite{jin2011, analytis2014, licciardello2019}.  Here, we present a high-field magnetotransport study on a series of Nd$_{1-x}$Sr$_x$NiO$_2$ (NSNO) thin films with $0.15 \leq x \leq 0.225$, spanning the entire SC dome. The moderate upper critical field \cite{li2019} $\mu_0 H_{\rm c2}$ $\lesssim$ 20~T in NSNO ensures that the normal state can be accessed in all SC films down to $T \approx$ 0.5~K under continuous magnetic fields up to 37.5~T.  At the lowest temperatures, the resistivity undergoes a logarithmic-divergence whose magnitude is systematically reduced with increasing doping. The resilience of the diverging resistivity to strong magnetic fields excludes localization or Kondo physics as its origin. Its resemblance to the logarithmically-divergent resistivity observed in underdoped cuprates \cite{ando1995, boebinger1996, ono2000} is noted, however, and suggests that the electronic ground state of infinite-layer nickelates is itself strongly correlated, though possibly not quantum critical. At $x = 0.225$, this insulating behavior cedes to a $T^2$ metallic resistivity characteristic of a correlated Fermi-liquid. Notably, no evidence of non-Fermi-liquid behavior associated with a putative quantum critical point was observed. 

%\section*{Methods}
Nd$_{1-x}$Sr$_x$NiO$_2$ thin films were grown by pulse laser deposition technique using the Ôhigh-fluenceÕ optimized growth conditions described in Ref. \cite{lee2020}. Electrical resistivity was measured using a four-point method with an alternating current $I$ = 10~$\mu$A at a frequency between 13 and 30~Hz, with the magnetic fields applied parallel to the crystalline $c$-axis. High-field measurements up to 37.5~T were performed using a Bitter magnet coupled to a $^3$He refrigerator at the High Field Magnet Laboratory in Nijmegen, the Netherlands.

\begin{figure*}[htp!!!]
\centering
\includegraphics[width=0.9\linewidth]{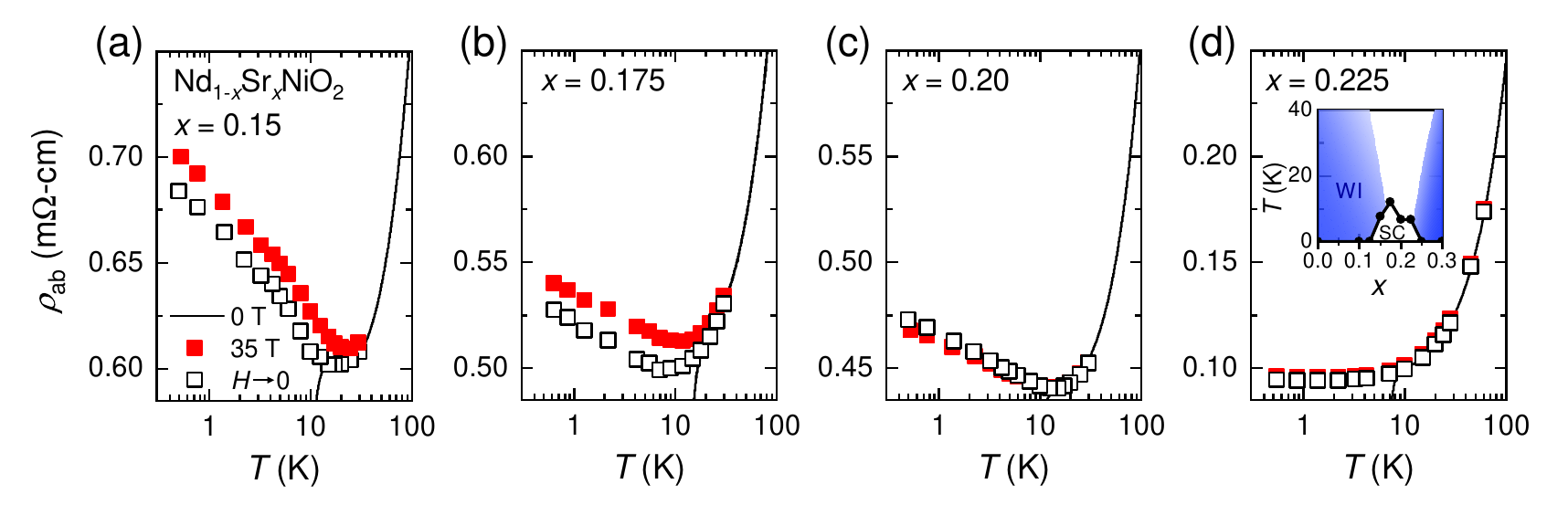}
\caption{Semi-log plots of the $T$-dependence of the in-plane resistivity $\rho_{ab}$ for (a) $x$ = 0.15 (b) $x$ = 0.175 (c) $x$ = 0.20 and (d) $x$ = 0.225 measured at zero magnetic field (solid line), at $\mu_0H$ = 35~T (filled squares), and extrapolations from the normal-state magnetoresistance to zero field (open squares). Error bars for $\rho_{ab}(H\rightarrow 0)$ are smaller than the symbols. The inset to panel (d) shows the $T-x$ phase diagram for Nd$_{1-x}$Sr$_x$NiO$_2$, with $T_{\rm c}$ defined as the mid-point of superconducting transition in the samples shown here. The blue shaded regions represent the weakly insulating (WI) behavior found on both sides of the superconducting (SC) dome \cite{li2020}. Note that the resistivity ranges in each panel are equivalent.}
\label{crossover}
\end{figure*}

\begin{figure}[h!]
\centering
\includegraphics[width=1\linewidth]{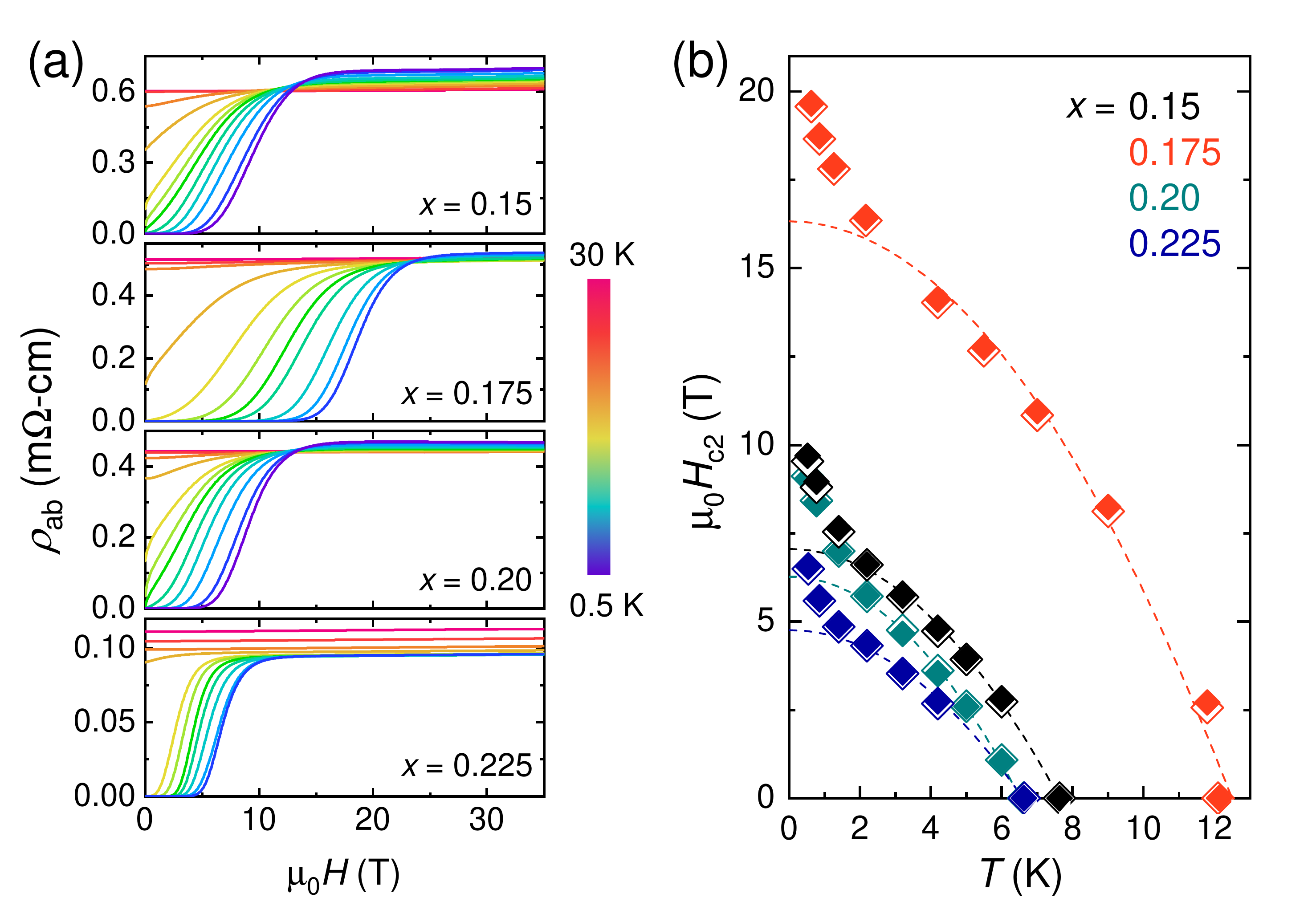}
\caption{(a) Resistivity isotherms as a function of magnetic field across the temperature range 0.5~K $\leq T \leq$ 30~K. 
(b) Upper critical field $H_{\rm c2}$ extracted using the 50\% $\rho_{\rm n}$ criterion (see main text for details). Open and filled symbols correspond to the extractions made using the extrapolated zero-field resistivity and the measured 35~T resistivity, respectively. Dashed lines are fits to the expression $\mu_0 H_{\rm c2}(T) = \mu_0 H_{\rm c2}(0)(1-(T/T_{\rm c})^2))$ down to 2~K, below which an anomalous upturn in $H_{\rm c2}$ is observed.}
\label{Hc2}
\end{figure}

%\section*{Result \& Discussion}
Figure~\ref{crossover}(a-d) shows the normal-state in-plane resistivity $\rho_{ab}(T)$ of four superconducting NSNO thin films with $x$ = 0.15, 0.175, 0.20 and 0.225, respectively. The zero-field normal-state resistivity is extrapolated by fitting the magnetoresistivity isotherms $\rho_{ab}(H)$ shown in Fig.~\ref{Hc2}(a) using the form of magnetoresistance (MR $\equiv \rho_{ab}(H) - \rho_{ab}(0)$) inferred from the field-derivatives (see details in the Supplemental Material \cite{SM1}). In short, we used a second-order polynomial function to fit the normal state resistivity curve in the high-field range where SC fluctuations are suppressed. At $T > T_{\rm c}$, the magnitude of the MR is small ($<$ 4\% at 35~T). Below $T_{\rm c}$, the MR remains small and, crucially, has a negligible $T$-dependence, thus the $T$-dependence of  $\rho_{ab}(H\rightarrow 0)$ and normal-state resistivity measured at sufficiently high fields are found to show the same qualitative behavior, independent of the strength of the magnetic field and/or assumed MR functional form (see Figs.~S2 and S3 in \cite{SM1}). Overall, our conclusions are thus independent of the details of the extrapolation methods.

For $x \neq$ 0.225, $\rho_{ab}(T)$ exhibits metallic behavior down to $T \approx$ 10~K before crossing over to insulating behavior with a log(1/$T$) divergence down to 0.5~K. The magnitude of the resistivity divergence diminishes systematically with increasing $x$ and is minimized at $x = 0.225$. A similar insulator-to-metal crossover was also observed in the field-induced normal state in various underdoped (UD) cuprates \cite{ando1995, boebinger1996, fournier1998, ono2000, li2002}. Indeed, the magnitude of the upturn in NSNO ($\sim$~10\%) is comparable to that observed in the electron-doped cuprates Pr$_{2-x}$Ce$_x$CuO$_4$ \cite{fournier1998} and Nd$_{2-x}$Ce$_x$CuO$_4$ \cite{li2002}, though it is around one order of magnitude smaller than the corresponding upturns in hole-doped La$_{2-x}$Sr$_x$CuO$_4$ \cite{ando1995, boebinger1996} and Bi$_2$Sr$_{2-x}$La$_x$CuO$_{6+\delta}$ \cite{ono2000}. The origin of the log($1/T$) resistivity in UD cuprates remains unclear. Nevertheless, it has been noted that low-$T$ upturns in $\rho_{ab}(T)$ are not universal in UD cuprates \cite{rullier-albenque2007, proust2016}, and irradiation studies have indicated that disorder, perhaps in combination with strong correlations, might play the dominant role \cite{rullier-albenque2008}. The resilience of the resistive upturn in NSNO against magnetic field as well as the positive MR (except for $x$ = 0.20) appears to rule out weak localization, Anderson localization and Kondo scattering \cite{zhang2020} as its origin. The systematic evolution of the upturn in NSNO with doping, coupled with the fact that for 0.15 $\leq x \leq$ 0.20, the resistive upturns begin either at or slightly below $T_c$, suggests some form of electronic order that competes with the superconductivity for spectral weight, such as the recently reported short-range antiferromagnetic correlations in NSNO \cite{lu2021}. While this conjecture also offers a viable explanation for the evolution of the metallic response (see below), it does not explain the re-entrant weakly insulating state beyond $x$ = 0.225.

To gain further insight into the low-$T$ electronic ground state of NSNO, we turn to examine the doping dependence of the resistively-determined upper critical field $H_{\rm c2}$. Here we use the criterion previously adopted \cite{wang2020} for NSNO, namely the field scale at which $\rho_{ab}(H)$ reaches 50\% of the normal-state resistivity $\rho_{\rm n}$. This choice is motivated by the reduced impact of SC fluctuations or vortex creep to its functional form \cite{brandow1998}, despite the fact that the SC order parameter remains finite at this field. Therefore, $H_{c2}$ shown in Fig.~\ref{Hc2} should be considered as the mean-field estimate and a lower bound. As can be seen, $H_{\rm c2}$ closely tracks $T_{\rm c}$ in NSNO, peaking at $\mu_0 H_{\rm c2} \approx$ 20~T for $x = 0.175$. For each value of $x$, $H_{\rm c2}$ also exhibits an anomalous upturn below $\approx$ 2~K, consistent with a previous finding \cite{wang2020} for $x = 0.225$, possibly indicating an unconventional or multi-gap SC ground state in the infinite-layer nickelates. We note that, while the choice of a different criterion e.g. $\rho_{\rm n, 90\%}$ or $\rho_{\rm n, 95\%}$ changes the magnitudes of $H_{\rm c2}$, the qualitative behavior of $H_{\rm c2}(T)$ and its evolution with doping remains the same (see Fig.~S5 in the Supplemental Material \cite{SM1}). Curiously, $H_{c2}$ is found to be the lowest for $x = 0.225$. This finding essentially excludes the possibility that the low-$T$ resistivity divergence becomes vanishingly small due to an incomplete suppression of the superconductivity. Rather, the doping-induced insulator-to-metal crossover is characteristic of the field-induced normal-state of the superconducting NSNO investigated here.

\begin{figure}[h!]
\centering
\includegraphics[width=1\linewidth]{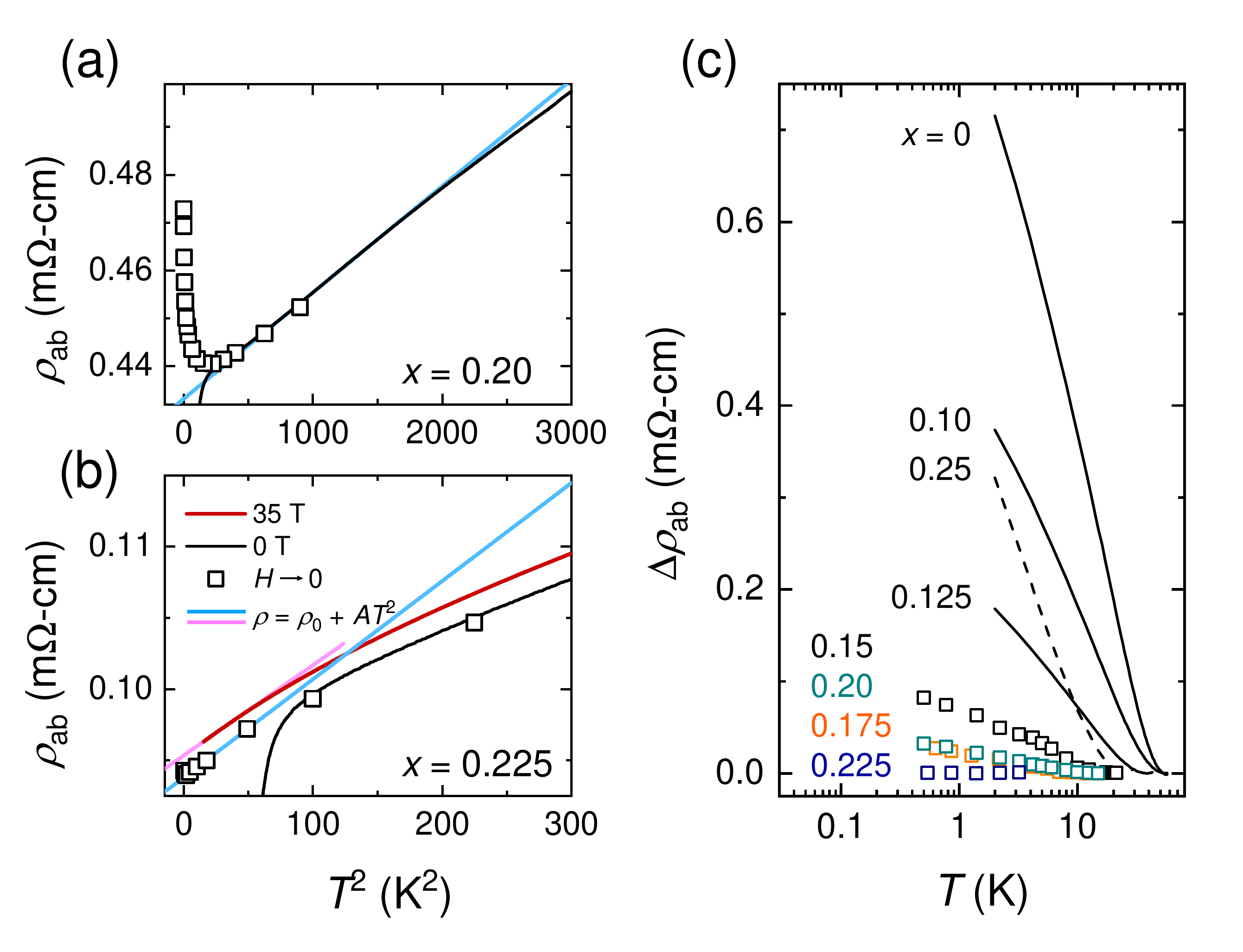}
\caption{(a,b) Zero-field resistivity as measured (black lines) and extrapolated (open squares) plotted as a function of $T^2$ for (a) $x$ = 0.20 and (b) $x$ = 0.225. Resistivity measured at 35~T down to 4.2~K is also shown for $x$ = 0.225 (red line). Blue and pink lines are fits made to zero-field and 35~T resistivity, respectively, using $\rho = \rho_0 + AT^2$. (c) Semi-log plot of the resistive upturn $\Delta\rho_{\rm ab} = \rho_{\rm ab} - \rho_{\rm min}$, after the resistivity value at its minimum $\rho_{\rm min}$ is subtracted from zero-field resistivity. Data of non-SC films are reproduced from \cite{li2020}.
}
\label{remnant}
\end{figure}

The suppression of the resistive upturn as $x \rightarrow 0.225$ provides an opportunity to examine more closely the form of low-$T$ metallic resistivity in NSNO. As shown in Fig.~\ref{remnant}, $\rho_{ab}$ for $x$ = 0.20 can be well-fitted to the expression $\rho = \rho_0 + AT^2$ for 300~K$^2 \lesssim T^2 \lesssim 2000$~K$^2$ (corresponding to 17~K $\lesssim T \lesssim$ 45~K) -- the lower bound being set by the onset of the resistive upturn. For $x$ = 0.225, $T^2$-resistivity is observed only below 7~K ($T^2 \lesssim$ 50~K). A pure $T^2$-resistivity as $T \rightarrow 0$ is one of the defining signatures of a Fermi-liquid ground state and its extension up to 45~K in Nd$_{0.8}$Sr$_{0.2}$NiO$_2$ suggests a predominance of electron-electron over electron-phonon scattering and the presence of strong (unscreened) electron correlations. The same analysis is applied to the other SC concentrations and summarized in Fig.~S6 and S7 in the Supplemental Material \cite{SM1}. While the $T^2$-regime is rather narrow for $x <$ 0.20, due to the growth of the resistivity upturn, we nevertheless find a systematic doping evolution of its onset temperature. To date, there is no doping level at which the low-$T$ resistivity is found to have a lower power-law exponent than 2 that might indicate proximity to a quantum critical point. Hence, while there is some evidence from the transport properties of a competing order, it does not, to date, exhibit clear signatures of quantum criticality.

The magnitude of the resistivity upturn $\Delta\rho_{ab}$, defined the difference between $\rho_{ab}(T)$ and the resistivity minimum $\rho_{\rm min}$, also exhibits a systematic doping evolution across the doping series. As shown in Fig.~\ref{remnant}(c), $\Delta\rho_{ab}$ is monotonically suppressed with increasing $x$ until $x = 0.225$, at which point, it becomes very small. For $x >$ 0.225, $\Delta\rho_{ab}$ again becomes sizeable. For lower-doped non-SC samples, the resistive upturn is closely correlated with an increase in the Hall coefficient $R_{\rm H}$ (compare Fig.~1(b) and Fig.~3(a) in Ref. \cite{li2020}), indicating that the upturn may be due to a drop in carrier density. In SC samples, the small upturns of order 5\% make it difficult to discern any change in $R_{\rm H}(T)$, particularly if $R_{\rm H}$ itself has an underlying $T$-dependence in the metallic state. In tandem with the systematic evolution of $\Delta\rho_{ab}(x)$, the drop in carrier density suggests the emergence of a second order parameter that competes with superconductivity for spectral weight, though this is not expected to impose a log$(1/T)$ dependence on the resistivity. We note that if the resistive upturn is indeed a signature of charge gapping, the gap, though small in magnitude, must be remarkably robust in the presence of an applied magnetic field. An alternative scenario for the upturn is the development of a secondary scattering mechanism. In this case, the upturn would be an additive contribution to $\rho_{ab}(T)$. In Fig.~S7 and Fig.~S8 in the Supplemental Material \cite{SM1}, we plot the remnant (insulating) contribution to $\rho_{ab}(T)$ after subtracting off a $T^2$-component, i.e. $\rho_{\rm rem} = \rho - \rho_{T^2}$ and a $T$-linear component, i.e. $\rho_{\rm rem} = \rho - \rho_{T}$, respectively. It is clear from Figs.~S7 and S8 that the systematic evolution of the resistive upturn is not qualitatively changed by our choice of fitting routine.

Curiously, while the metallicity is optimized at $x = 0.225$, the strength of the superconductivity in NSNO, as reflected in the absolute magnitude of $H_{\rm c2}$, is not. In fact, the $T_{\rm c}$ and $H_{\rm c2}$ are highest at $x = 0.175$, showing no clear correlation to the normal state properties in the series of samples investigated here. Evidently, the relation between normal state and superconducting state properties in unconventional superconductors is a complex one, with multiple instabilities vying for spectral weight and frustrating the outcome for the eventual dominant ground state.

%\section*{Conclusion}
In summary, we find a number of salient features of the normal electronic ground state of superconducting nickelates. Firstly, the weakly insulating behavior observed outside the SC dome is found to persist in the field-induced normal state within the SC regime. Secondly, at all dopings except $x = 0.225$, the low-$T$ resistivity follows closely a log(1/$T$) divergence, behavior reminiscent of that found in various underdoped cuprates \cite{ando1995, boebinger1996, fournier1998, ono2000, li2002}. Thirdly, prior to the onset of the upturn, the $T$-dependence of the normal-state resistivity in the metallic regime undergoes a crossover from $T$-linear at high-$T$ to $T$-quadratic at lower temperatures.  Lastly, at $x$ = 0.225 where the resistivity divergence is the smallest, the resistivity below 10~K can be well-described by the expression $\rho = \rho_0 + AT^2$, suggestive of the recovery of a correlated Fermi-liquid state. The resilience of the resistivity upturn against magnetic fields appears to preclude localization or Kondo scattering as the origin of the non-metallic behavior. Rather, it implies the presence of strong unscreened electron correlations, possibly associated with fluctuations of a competing order parameter.

%\section*{Acknowledgement}
We acknowledge the support of the HFML-RU/NWO, a member of the European Magnetic Field Laboratory (EMFL). This work was supported by the Netherlands Organisation for Scientific Research (NWO) grant No. 16METL01 \lq Strange Metals' and the European Research Council (ERC) under the European Union's Horizon 2020 research and innovation programme (Grant Agreement No. 835279-Catch-22). The work at SLAC/Stanford is supported by the Gordon and Betty Moore Foundations Emergent Phenomena in Quantum Systems Initiative through grant number GBMF9072 (synthesis equipment) and the US Department of Energy, Office of Basic Energy Sciences, Division of Materials Sciences and Engineering, under contract number DE-AC02-76SF00515.

\bibliographystyle{apsrev4-2}

\end{document}